\documentclass[pra,twocolumn,showpacs,superscriptaddress]{revtex4-1}
\usepackage{graphicx}
\usepackage{amssymb}
\usepackage{amsfonts}
\usepackage{amsmath}
\usepackage{lmodern}
\usepackage{mathrsfs}
\usepackage[dvipdfm,colorlinks=true, citecolor=blue, urlcolor=blue ]{hyperref}
\hypersetup{breaklinks=true}

\begin{document}

\title{Anomalous decoherence in a dissipative two-level system }
\author{Hai-Bin Liu}
\affiliation{Center for Interdisciplinary Studies $\&$ Key Laboratory for Magnetism and Magnetic Materials of the MoE, Lanzhou University, Lanzhou 730000, China}
\author{Jun-Hong An}
\email{anjhong@lzu.edu.cn}
\affiliation{Center for Interdisciplinary Studies $\&$ Key Laboratory for Magnetism and Magnetic Materials of the MoE, Lanzhou University, Lanzhou 730000, China}
\affiliation{Center for Quantum Technologies and Physics Department, National University of Singapore, 3 Science Drive 2, Singapore 117543, Singapore}
\author{Chong Chen}
\affiliation{Center for Interdisciplinary Studies $\&$ Key Laboratory for Magnetism and Magnetic Materials of the MoE, Lanzhou University, Lanzhou 730000, China}
\author{Qing-Jun Tong}
\affiliation{Center for Interdisciplinary Studies $\&$ Key Laboratory for Magnetism and Magnetic Materials of the MoE, Lanzhou University, Lanzhou 730000, China}
\author{Hong-Gang Luo}
\affiliation{Center for Interdisciplinary Studies $\&$ Key Laboratory for Magnetism and Magnetic Materials of the MoE, Lanzhou University, Lanzhou 730000, China}
\affiliation{Beijing Computational Science Research Center, Beijing 100084, China}
\author{C. H. Oh}
\email{phyohch@nus.edu.sg}
\affiliation{Center for Quantum Technologies and Physics Department, National University of Singapore, 3 Science Drive 2, Singapore 117543, Singapore}

\begin{abstract}
We study systematically the non-Markovian decoherence dynamics of a dissipative two-level system, i.e., the so-called spin-boson model. It is interesting to find that the decoherence tends to be inhibited with the increase of the coupling strength between the system and the reservoir, which is contrary to the common recognition that a stronger coupling always induces a severer decoherence. This is attributed to the occurrence of a quantum phase transition (QPT). The relationship between this QPT and conventional delocalized-localized QPT is also discussed. Our result suggests a useful control method to overcome the detrimental effects of the reservoir to the system.
\end{abstract}
\maketitle

\section{Introduction}
The decoherence of a dissipative two-level system (TLS) as a qubit, described by the spin-boson model (SBM), is a main obstacle to the practical realization of quantum information processing \cite{Nielsen2000}. Recently, the quantum phase transition (QPT), which describes a sudden qualitative change of the macroscopic properties mapped from the eigenspectrum of a quantum many-body system \cite{Sachdev1999}, of the SBM also is attracting much attention. One of the motivations is that people hope that the exploration of the QPT in the SBM will supply some insight into the decoherence control of the qubit system.
Due to its unsolvability, a variety of approximate analytical and numerical methods, for example, the path-integral method under a noninteracting blip approximation \cite{Leggett1987}, variational method based on unitary transformation \cite{Silbey1984,Zheng2007,Chin2007}, numerical renormalization-group method \cite{Bulla2003,Vojta2012}, quantum Monte Carlo method \cite{Winter2009}, and numerical diagonalization in a coherent-state basis \cite{Zhang2010}, have been developed. A consensus is that the SBM shows a QPT from delocalization to localization with the increase of the coupling strength in the case of Ohmic spectral density, as a consequence of the competition between the internal transition effect of the TLS and the external dissipation effect of the reservoir.

Compared with the Ohmic case, the QPT of the SBM with sub-Ohmic spectral density, which has been used to model the $1/f$ noise \cite{Paladino2002} in quantum dots \cite{Jung2004,Tobias2008} and superconductor qubit systems \cite{Astafiev2006,Eroms2006}, is more involved. Different methods from the sub-Ohmic SBM cannot even lead to a qualitatively consistent result. The path-integral method under a noninteracting blip approximation predicts that the QPT from delocalization to localization is absent for the sub-Ohmic SBM \cite{Leggett1987}. The numerical renormalization-group method confirms the occurrence of the QPT in the full range $0<s<1$ (here, $s$ is the exponent of the spectral density), while the breakdown of the quantum-to-classical mapping for $0<s<1/2$ \cite{Vojta2005}, which means the failure of the classical mean-field description to the QPT. However, the quantum Monte Carlo method and the numerical diagonalization in the coherent-state basis predict the presence of the QPT, the well-defined quantum-to-classical mapping, and the classical critical exponents to the sub-Ohmic SBM \cite{Winter2009,Zhang2010}.

Due to the rich physics in the SBM, the decoherence dynamics of the TLS also shows rich behaviors. It was found that the decoherence dynamics of the TLS of the SBM under the rotating-wave approximation (RWA) exhibits the exponential decay under the Born-Markovian approximation \cite{Breuer2002}, the oscillatory decay in a lossy cavity with Lorentzian spectral density \cite{Rempe1987}, and even the decoherence suppression in the engineered reservoir with photonic band-gap structure \cite{Yablonovitch1987,John1994,Tong3}. If the RWA is relaxed, it was shown that the dynamics of the spin in the delocalized phase regime changes from the damped coherent oscillation to incoherent relaxation with the increase of the coupling strength for both the Ohmic \cite{Leggett1987} and the sub-Ohmic \cite{Chin2006,Anders2007,Wang2009} SBM. This dynamical phenomenon is named the coherent-incoherent transition \cite{Leggett1987}. What is the physical reason for such dynamical transition and these rich dynamical behaviors?

In the present work, we study the non-Markovian dynamics of the TLS of the SBM, both with and without the rotating-wave approximation. Under the RWA, we find analytically that the decoherence tends to be inhibited with the increase of the coupling constant between the system and the reservoir, which is contrary to the common recognition that a stronger coupling always induces a severer decoherence. We also show that this  anomalous behavior is caused by the occurrence of an intrinsic quantum phase transition (QPT) of the SBM. When the RWA is relaxed, using the perturbation approach based on a unitary transformation, which has been successfully used to capture the delocalized-localized QPT for the Ohmic \cite{Silbey1984,Zheng2007,Tong2011} and the sub-Ohmic \cite{Wang2009} SBM, we find that the QPT that happened explicitly in the conventional delocalized phase regime still exists. The qualitative compatibility of this QPT with the coherent-incoherent transition \cite{Kast2013} makes us conjecture that the coherent-incoherent transition occurring in the delocalized regime is actually caused by an intrinsic QPT. Our analytical formulation provides a clear physical picture of this QPT and a unified description of the QPT in the sub-Ohmic SBM.

This paper is organized as follows. In Sec. \ref{model}, the SBM and its simplification under the RWA are introduced. The anomalous decoherence of the SBM under the RWA and its essential reason, i.e., the QPT, is revealed in Sec. \ref{bdt} by examining the formation of a bound state. In Sec. \ref{qptsb}, the anomalous decoherence and the QPT of the SBM without the RWA is investigated by means of the perturbation approach based on a unitary transformation \cite{Silbey1984,Zheng2007}. Finally, a brief discussion and summary are given in Sec. \ref{sum}.

\section{The model}\label{model}
The SBM, which relates to a variety of physical and chemical processes \cite{Weiss}, describes the interaction of an effective TLS with a bosonic reservoir. Its Hamiltonian reads
\begin{equation}
\hat{H}={\epsilon\over 2}\hat{\sigma}_z-\frac{\Delta}{2}\hat{\sigma} _{x}+\sum_{k}\omega
_{k}\hat{b}_{k}^{\dagger }\hat{b}_{k}+\sum_{k}\frac{g_{k}}{2}\hat{\sigma}
_{z}(\hat{b}_{k}+\hat{b}_{k}^{\dagger }),  \label{sbm}
\end{equation}
where $\epsilon$ and $\Delta$, respectively, are the energy difference and the transition amplitude between the two levels, and $\hat{b}_{k}^{\dagger }$ ($\hat{b}_{k}$) is the creation (annihilation) operator of the $k$th mode of the reservoir with frequency $\omega _{k}$. The coupling strength between the TLS and its reservoir is denoted by $g_{k}$, which is further characterized by the spectral density $J(\omega )=\pi\sum_{k}\left\vert {g_{k}}\right\vert^{2}\delta(\omega-\omega_{k})$. In the continuum limit, the spectral density may have the form
\begin{equation}J(\omega)=2\pi\alpha\omega_c^{1-s}\omega^s\Theta(\omega_c-\omega),\label{sp}
\end{equation}
where $\alpha$ is a dimensionless coupling constant, $\omega_c$ is a cutoff frequency, and $\Theta(x)$ is the usual step function. The reservoir is classified as Ohmic when $s = 1$, sub-Ohmic when $0 < s < 1$, and super-Ohmic when $s > 1$ \cite{Leggett1987}. In spite of the simplicity of its formulation, the SBM does not admit an exact solution in a closed analytical form and one often resorts to numerical simulations or various approximations for its analysis. Under a unitary transformation $\hat{U}_{1}=\exp(-i\pi\hat{\sigma}_{y}/4)$, one can prove that Eq. (\ref{sbm}) is equivalent to
\begin{equation}
\hat{H}_z={\epsilon\over 2}\hat{\sigma}_x+\frac{\Delta}{2}\hat{\sigma} _{z}+\sum_{k}\omega
_{k}\hat{b}_{k}^{\dagger }\hat{b}_{k}+\sum_{k}\frac{g_{k}}{2}\hat{\sigma}_{x}(\hat{b}_{k}+\hat{b}_{k}^{\dagger }),
\label{uth}
\end{equation}
which corresponds to a $\pi/ 4$ rotation around $\hat{\sigma}_y$ to Eq. (\ref{sbm}). In the following, we assume $\epsilon=0$ for simplicity.

The interaction in Eq. (\ref{uth}) contains the counter-rotating terms, $\hat{b}_k^\dag\hat{\sigma}_+$ and $\hat{b}_k\hat{\sigma}_-$. A widely used approximation in quantum optics and quantum information communities is the RWA, which is applicable in the weak-coupling limit. Then, Eq. (\ref{uth}) is reduced to
\begin{equation}
\hat{H}_{\text{RWA}}={\Delta\over 2}\hat{\sigma} _{z}+\sum_{k}\omega _{k}\hat{b}_{k}^{\dagger
}\hat{b}_{k}+\sum_{k}{g_{k}\over 2}(\hat{\sigma} _{+}\hat{b}_{k}+\hat{\sigma} _{-}\hat{b}_{k}^{\dagger
}),  \label{rwa}
\end{equation}
which is analytically solvable because the total excitation number $\hat{N}=\sum_k\hat{b}_k^\dag \hat{b}_k+\hat{\sigma}_+\hat{\sigma}_-$ of the whole system is conserved.

\section{Decoherence inhabitation under RWA}\label{bdt}

The Hamiltonian (\ref{rwa}) under the RWA is widely used to characterize decoherence of a qubit in quantum optics and quantum information communities \cite{Nielsen2000}. In this section, starting from Eq. (\ref{rwa}), we investigate analytically and numerically the decoherence dynamics of the TLS and reveal that an anomalous decoherence phenomenon, i.e., decoherence inhabitation, exists in the model. It is essentially caused by an intrinsic QPT of the whole system.
\subsection{Decoherence dynamics}
Assume initially that the whole system is in
\begin{equation}|\Psi_1(0)\rangle=|+,\{0_k\}\rangle,\label{psi1}
\end{equation}which, in the $\hat{H}_z$ [Eq. (\ref{uth})] representation, takes the form $|\Psi_z(0)\rangle=\hat{U}_1|\Psi_1(0)\rangle=|+_x,\{0_k\}\rangle$, with $|+_x\rangle$ satisfying $\hat{\sigma}_x|+_x\rangle=|+_x\rangle$. The time evolution of $|\Psi_z(0)\rangle$ is governed by Eq. (\ref{rwa}) under the RWA. Therefore, the time-dependent solution can be expanded as
\begin{equation}|\Psi_z(t)\rangle=e^{{i\Delta t\over2}}[{1\over\sqrt{2}}|-,{0_k}\rangle+c(t)|+,{0_k}\rangle+\sum_k d_k(t)|-,1_k\rangle].
\end{equation}From the Schr\"{o}dinger equation, we can get the probability amplitude $c(t)$ satisfying
\begin{equation}\dot{c}(t)+i\Delta c(t)+\int_0^t f(t-\tau)c(\tau)d\tau=0,
\end{equation}where the kernel function $f(t-\tau)\equiv {1\over 4\pi}\int_0^{+\infty} J(\omega)e^{-i\omega(t-\tau)}d\omega$ and the initial condition $c(0)=1/\sqrt{2}$. With this result in hand, we can calculate $P_z\equiv\langle\Psi(0)| e^{i\hat{H}t}\hat\sigma_ze^{-i\hat{H}t}|\psi(0)\rangle$ under the RWA as
\begin{equation}P_{z}(t)=\langle\Psi_z(t)|\hat{\sigma}_x|\Psi_z(t)\rangle=\sqrt{2}\text{Re}[c(t)].\label{pz1}
\end{equation}

\begin{figure}[h]
\centering
\includegraphics[width = 1 \columnwidth]{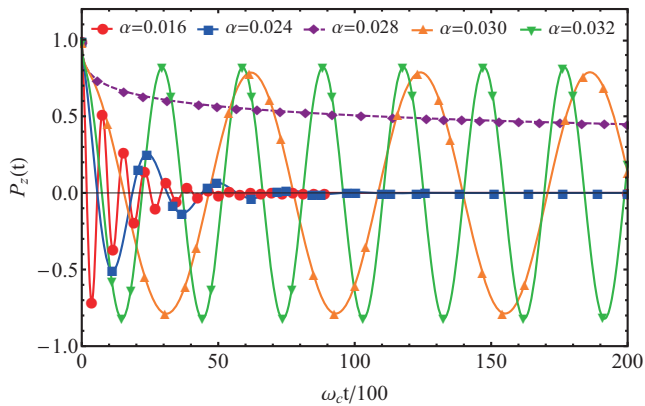}
\caption{(Color online) The nonequilibrium dynamics of $P_z(t)$ of the SBM with the RWA under the initial condition given by Eq. (\ref{psi1}). Here the parameters $s=0.7$ and $\Delta=0.02\omega_c$ have been used. Consequently, the critical point is $\alpha_{_\text{C,RWA}}=0.028$.  } \label{dyn1}
\end{figure}

In Fig. \ref{dyn1}, we plot $P_z(t)$ of Eq. (\ref{pz1}) in different coupling constants. We can see that $P_z(t)$ decays to zero after transient oscillations in the weak coupling limit. We refer to the character of the dynamics in this region as the complete decoherence. With the increasing of the coupling constant, the decay is surprisingly suppressed, which is dramatically different from one's expectation that a stronger coupling always induces a severer decoherence. When $\alpha=0.028$, the coherence does not decay to zero and a finite quantum coherence is preserved in the steady state. With the further increasing of $\alpha$, the oscillation is totally stabilized and the decoherence is inhibited. We call the character of the dynamics in the region $\alpha\geq \alpha_{_\text{C,RWA}}$ as the decoherence suppression.

\subsection{QPT under the RWA}
The physical mechanism of this anomalous decoherence behavior can be understood from the eigenspectrum of the whole system. Since $\hat{N}$ is conserved, the Hilbert space is split into the direct sum of the subspaces with definite quantum number $N$. In this situation, one can naively deem that the eigenstate $|\varphi_0\rangle=\left\vert-,\{0_k\}\right\rangle$, i.e., a tensor product of the respective ground states of the two subsystems in zero-excitation subspace with eigenvalue $E_0=-\Delta/ 2$ is the ground state of the whole system. Is this always true? To verify this, let us examine the eigensolution of $\hat{H}_\text{RWA}$ in the single-excitation subspace, which can be expanded as $\left\vert \varphi _{1}\right\rangle =c_{0}\left\vert
+,\{0_{k}\}\right\rangle +\sum_{k=0}^{\infty }c_{k}\left\vert-,1_{k}\right\rangle $. From the eigenequation governed by Eq. (\ref{rwa}), we can obtain a transcendental equation of $E_1$,
\begin{equation}
y(E_1)\equiv {\Delta\over 2}- {1\over 4\pi}\int_{0}^{\infty }\frac{%
J(\omega) }{\omega -(E_1+{\Delta\over 2})}d\omega =E_1.  \label{t11}
\end{equation}
A bound state is an eigenstate with real (discrete) eigenvalue in a quantum many-body system. So if Eq. (\ref{t11}) has a real root, then we can claim that the system possesses a bound state \cite{John1990,Nakazato1996,Miyamoto05,Tong1}. We can easily find that $y(E_1)$ decreases monotonically with the increase of $E_1$ in the regime of $E_1<-{\Delta\over 2}$. Therefore, if the condition
\begin{equation}
y(-{\Delta\over 2})\leq-{\Delta \over 2},\label{cr}
\end{equation}
is satisfied, then $y(E_1)$ always has one and only one intersection with the function on the right-hand side of Eq. (\ref{t11}). This root just corresponds to the eigenvalue of the formed bound state in the Hilbert space of the system plus its reservoir. On the other hand, in the regime of $E_1>-{\Delta\over 2}$, we can see that $y(E_1)$ is divergent, which means that no real root $E_1$ can make Eq. (\ref{t11}) well defined. Consequently, Eq. (\ref{t11}) does not have a real root to support the existence of a further bound state in this regime. The excited-state population in the bound state, as a stationary state of the whole system, is constant in time. This means that the formation of the bound state can result in decoherence suppression.

The formation of a bound state in the SBM is reminiscent of the study in the Friedrichs model \cite{Rzazewski1982,Lewenstein2000,Miyamoto05}, where a similar bound state was revealed and the corresponding dynamics was studied. Here we argue further that accompanying the ground state changing from $|\varphi_0\rangle$ to the bound state $|\varphi_1\rangle$ (due to $E_1<E_0$), the formation of the bound state actually corresponds to a quantum phase transition. We can verify that the eigenstates of $\hat{H}_\text{RWA}$ in the subspaces $N\geq2$ actually have larger eigenvalues than $E_1$. This implies that the higher-boson states may not become the ground state. One may also observe that the two states are orthogonal, i.e., $\left\langle-,\{0_{k}\}\vert\varphi_1\right\rangle=0$. Therefore, the energy-level crossing accompanying the formation of the bound state signals clearly that the system undergoes a QPT. From the criterion (\ref{cr}), it is straightforward to evaluate that the QPT happens at the critical point
\begin{equation}\label{cp}
    \alpha_{_{\text{C,RWA}}}={2 s\Delta \over \omega_c},
\end{equation}for the spectral density (\ref{sp}).
Experimentally, the bound-state-induced decoherence suppression has been observed \cite{Lodahl2004,Noda2007,Dreisow08}.

\begin{figure}[tbp]
\centering
\includegraphics[width = 1 \columnwidth]{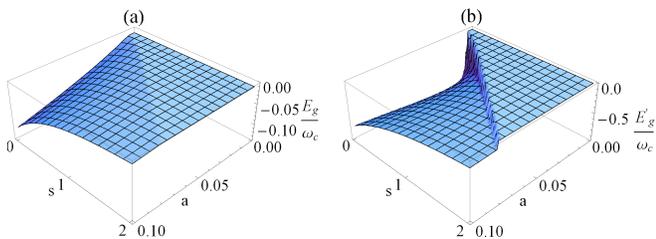}
\caption{(Color online) (a) Ground-state energy $E_g$ and (b) its first derivative $E_g'={\partial E_g\over \partial \alpha}$ as a function of the coupling constant $\alpha$ and power index $s$ of the spectrum. The parameter used here is $\Delta=0.02\omega_c$. According to Eq. (\ref{cp}), the first-order QPT occurs at $\alpha_{_{\text{C,RWA}}}=0.04s$, which has been confirmed by the discontinuousness of the first derivative of the ground-state energy.}
\label{fig2}
\end{figure}
\begin{figure}[tbp]
\centering
\includegraphics[width = 1 \columnwidth]{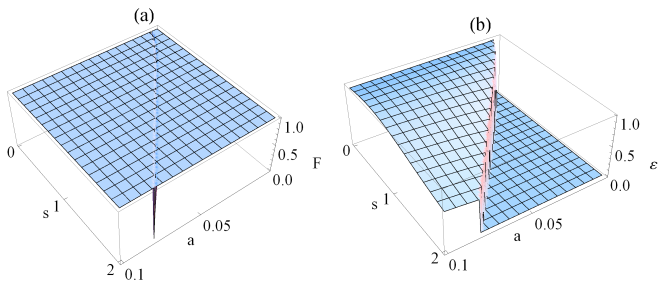}
\caption{(Color online) (a) Ground state fidelity and (b) entanglement entropy as a function of the coupling constant $\alpha$ and power index $s$ of the spectrum. Here, $\delta\alpha=0.0005$; other parameters used here are the same as Fig. \ref{fig2}. The singularity near the critical point $\alpha_{_{\text{C,RWA}}}$ of forming the bound state shows the existence of QPT in this model.}
\label{fig3}
\end{figure}
To verify the existence of QPT in the model quantitatively, in the following we study numerically the ground-state energy and its derivative, fidelity and entanglement entropy of the ground state near the critical point with the change of the coupling constant of the system.

At zero temperature, the nonanalyticity of the ground-state energy is directly connected to the QPT. The first (or $n$th) order QPT is characterized by the discontinuity in the first (or $n$th) derivative of the ground-state energy. In Fig. \ref{fig2}, we plot the ground-state energy and its first derivative. It can be seen that the first derivative is discontinuous at the critical point (\ref{cp}), which means that it is a first-order QPT.

The QPT can be further verified by the fidelity $F$ and entanglement entropy $\mathcal {E}$ between the TLS and the reservoir of the ground state. The ground-state fidelity is defined as the overlap of two ground states corresponding to two slightly different control parameters \cite{Zanardi06},
i.e., $F=\langle \varphi_g(\alpha)|\varphi_g(\varphi+\delta\alpha)\rangle$. The entanglement entropy can be obtained by calculating the entropy of the reduced density matrix of the TLS after tracing out the reservoir degrees of freedom. In Fig. \ref{fig3}(a), we plot $F$ near the critical point. The singularity in the plot evidences clearly the existence of QPT in this model. Because of the totally orthogonal property of the ground state, the fidelity completely drops to zero at the critical point. In Fig. \ref{fig3}(b), we plot $\mathcal {E}$ of the ground state. Near the critical point, we find a sudden birth of the ground-state entanglement, which can be seen as a result of the changing of the ground-state structure. This discontinuity in the ground-state entanglement entropy also evidences the existence of QPT.

With this QPT in hand, the dynamics in Fig. \ref{dyn1} can be easily understood. According to Eq. (\ref{cp}), the QPT for the parameters there occurs at $\alpha_{_\text{C,RWA}}=0.028$. If $\alpha<\alpha_{_\text{C,RWA}}$, then the dynamics reduces the oscillatory damping to zero. It is understandable from the fact that the bound state in this region is absent and all of the quantum coherence decays to zero. When $\alpha=\alpha_{_\text{C,RWA}}$, the bound state $|\varphi_1\rangle$ with the eigenvalue being just $E_1=-\Delta/2$, which is equal to $E_0$, is formed. As a stationary state, the quantum coherence contributed from the component $|\varphi_1\rangle$ to Eq. (\ref{psi1}) does not change during the time evolution. Therefore, we get a finite asymptotical $P_z(t)$. With the further increase of $\alpha$, the dynamics shows lossless oscillation. The bound state with smaller eigenvalue than $E_0$ is present. In this region, the components of $|\varphi_0\rangle$ and $|\varphi_1\rangle$ in Eq. (\ref{psi1}) have different time dependence. The difference of the two eigenvalues, i.e., $E_0-E_1$, contributes to the frequency of this lossless oscillation. A larger $\alpha$ induces a smaller $E_1$ and a faster oscillation of $P_z(t)$. All of this analysis matches well with the numerical results in Fig. \ref{dyn1}.

\section{The anomalous decoherence without the RWA}\label{qptsb}
In the following, we evaluate the correctness of the counter-rotating terms to the dynamics and the QPT. We first recover the conventional delocalized-localized QPT. Then, using the perturbation approach based on unitary transformation \cite{Silbey1984,Zheng2007}, we study the anomalous decoherence in the delocalized phase regime and its intrinsic mechanism, i.e., the existence of a further QPT in this conventional phase regime.

\subsection{The conventional delocalized-localized QPT}
A unitary transformation $\hat{U}_2=\exp[\sum_{k}\frac{g _{k}\xi_{k}}{2\omega_{k}}(\hat{b}_{k}^{\dagger}-\hat{b}_{k})\hat{\sigma} _{x}]$ can recast Eq. (\ref{uth}) into $\hat{H}^{\prime }=\hat{H}_{0}^{\prime }+\hat{V}^{\prime }$, where
\begin{eqnarray}
\hat{H}_{0}^{\prime } &=&\frac{\eta \Delta }{2}\hat{\sigma} _{z}+\sum_{k}\omega_{k}\hat{b}_{k}^{\dag }\hat{b}_{k}+C,\nonumber \\
\hat{V}^{\prime } &=&\sum_{k}\frac{g_{k}(1-\xi _{k})}{2}(\hat{b}_{k}+\hat{b}_{k}^{\dag})\hat{\sigma} _{x}-i\frac{\Delta}{2}\hat{\sigma} _{y}\sinh\hat{\chi} \nonumber\\
&&+\frac{\Delta }{2}\hat{\sigma} _{z}(\cosh \hat{\chi} -\eta ),\label{ff}
\end{eqnarray}%
with $C=\sum_{k}\frac{g_{k}^{2}}{4\omega _{k}}\xi_{k}(\xi _{k}-2)$, $\hat{\chi} =\sum_{k}\frac{g_{k}\xi _{k}}{\omega _{k}}(\hat{b}_{k}^{\dag }-\hat{b}_{k})$, and \begin{equation}\eta =\langle \{0_{k}\}|\cosh \hat{\chi} |\{0_{k}\}\rangle =\exp [-\sum_{k}\frac{g_{k}^{2}\xi _{k}^{2}}{2\omega _{k}^{2}}].\label{eta1}\end{equation} We can see from Eqs. (\ref{ff}) that, to the zero-order approximation, the spin-boson interactions can be eliminated to generate an effective noninteracting Hamiltonian characterized by a renormalized transition amplitude $\Delta _{eff}\equiv \eta \Delta $, with $\eta$ as the renormalized factor.

With the separation of Eqs. (\ref{ff}), we can readily calculate the Bogoliubov-Peierls bound on the free energy $F_{B}$ of the system \cite{Feynman}. The free energy $F$ of the system is related to $F_{B}$ by $F \leq F_{B}$ with
\begin{eqnarray}
F_{B} =-\beta ^{-1}\ln \text{Tr}\exp (-\beta \hat{H}_{0}^{\prime })+\langle \hat{V}^{\prime }\rangle _{\hat{H}_{0}},
\end{eqnarray}%
where $\beta =\frac{1}{k_{B}T}$, $\langle \cdot \rangle _{\hat{H}_{0}^{\prime }}$ denotes the thermal expectation value calculated with respect to
$\hat{H}_{0}^{\prime }$, and the trace is calculated using the eigenstates of $\hat{H}_{0}^{\prime }$. It is easy to find $\langle \hat{V}^{\prime }\rangle_{H_{0}}=0 $. Therefore,
\begin{equation}
F_{B}=F_{\text{Boson}}-{\ln [2\cosh \frac{\beta \Delta \eta }{2}]\over \beta}+C.
\end{equation}%
The parameters $\xi _{k}$ are determined by minimizing $F_{B}$ with respect to $\xi _{k}$, that is, $\frac{\partial F_{B}}{\partial \xi _{k}}=0$. We find, in our zero-temperature case (i.e., $\beta \rightarrow \infty$),
\begin{equation}
\xi _{k}=\frac{\omega _{k}}{\omega _{k}+\eta \Delta }.\label{xi}
\end{equation}

By now, the parameters $\xi_k$ as well as the renormalized factor $\eta$ have been determined. The renormalized factor $\eta$ has been used successfully to characterize the delocalized-localized QPT in the SBM \cite{Silbey1984, Zheng2007, Wang2009}. If the transition amplitude is renormalized to zero, then the system is in the localized phase and the dynamics is trivial. In contrast, if the renormalized transition amplitude is nonzero, then the system is in the delocalized phase.
\begin{figure}[h]
\centering
\includegraphics[width = 0.9 \columnwidth]{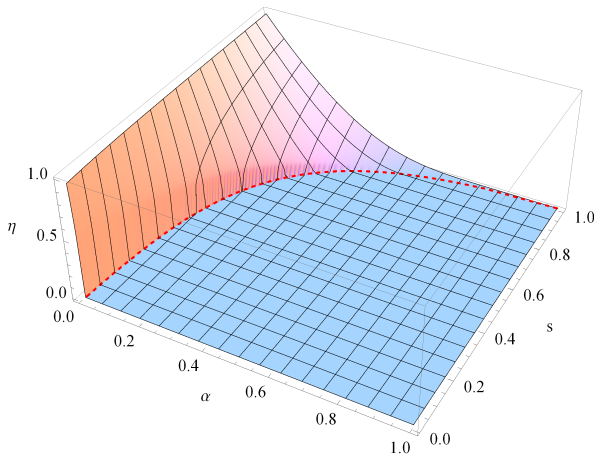}
\caption{(Color online) The conventional delocalized-localized QPT characterized by the renormalized factor $\eta$ as a function of the coupling constant $\alpha$ and the power index $s$ of the spectrum. The red dashed line depicts the critical point. $\Delta=0.02\omega_c$ has been used in the numerical calculation.} \label{etadl}
\end{figure}

In Fig. \ref{etadl}, we plot the numerical results on this conventional QPT characterized by $\eta$, which can be calculated by solving Eqs. (\ref{eta1}) and (\ref{xi}) self-consistently. We can see that the system is in the delocalized phase regime when $\alpha$ is small, where $\eta$ takes a finite value. With the increase of $\alpha$, $\eta$ drops suddenly to zero and the system enters into the localized phase regime. Such delocalized-localized QPT is present in the whole range of the power index $s$ of the sub-Ohmic spectral density. This is coincident with the results under the quantum Monte Carlo method and the numerical diagonalization method \cite{Winter2009,Zhang2010}. The critical point of this QPT tends to $\alpha_{_\text{C}}=1$ with the increase of $s$ to the Ohmic case. This is consistent with the well-known results that the delocalized-localized phase transition occurs at $\alpha_{_\text{C}}=1$ in the small $\Delta$ limit for the Ohmic SBM \cite{Leggett1987}.

\subsection{The anomalous decoherence in the delocalized phase regime}

Focusing on the delocalized phase regime, which occurs in the weak coupling limit, we further separate the first-order perturbation term from $\hat{V}^{\prime }=\hat{H}_{1}^{\prime }+\hat{H}_{2}^{\prime }$ with
\begin{eqnarray}
\hat{H}_{1}^{\prime }
&=&\sum_{k}\nu _{k}(\hat{b}_{k}\hat{\sigma} _{+}+\hat{b}_{k}^{\dag }\hat{\sigma} _{-}),  \nonumber \\
\hat{H}_{2}^{\prime } &=&\frac{\Delta }{2}\hat{\sigma} _{z}(\cosh \hat{\chi} -\eta )-i\frac{\Delta }{2}\hat{\sigma} _{y}(\sinh \hat{\chi} -\eta \hat{\chi} ),  \label{vv}
\end{eqnarray}%
where $\nu _{k}=\eta \Delta g_{k}\xi _{k}/\omega _{k}$ and $\hat{\sigma}_{\pm }=(\sigma _{x}\pm i\sigma _{y})/2$. Combining with Eq. (\ref{ff}),
we arrive at the transformed SBM as $\hat{H}^{\prime }=\hat{H}_{0}^{\prime }+\hat{H}_{1}^{\prime }+\hat{H}_{2}^{\prime }$, where $\hat{H}_{0}^{\prime }$ collects all the renormalized noninteracting terms, $\hat{H}_{1}^{\prime }$ collects all the first-order perturbation terms, and $\hat{H}_{2}^{\prime }$ collects all the higher-order ones. It has been proved that in zero-temperature and weak-coupling (i.e., the delocalized) regimes, the higher-order perturbation terms $\hat{H}_2'$ can be neglected \cite{Zheng2007}, which also can be proved self-consistently by our following results. Then the transformed Hamiltonian has the form $\hat{H}'\approx \hat{H}'_{0}+\hat{H}'_{1}\equiv \hat{H}_\text{eff} $,
\begin{eqnarray}
\hat{H}_\text{eff}=\frac{\Delta\eta }{2}\hat{\sigma} _{z}+ \sum_{k}[\omega_{k}\hat{b}_{k}^{\dag}\hat{b}_{k}+\nu_{k}(\hat{b}_{k}^{\dag}\hat{\sigma}_{-}+\hat{b}_{k}\hat{\sigma}_{+})]+C,\label{rwa22}
\end{eqnarray}
which shares the formal similarity with the rotating-wave approximate Hamiltonian (\ref{rwa}).

Next, we study the dynamics when the RWA is relaxed. To make the dynamics manifest the effect of the formed bound state exclusively, we choose the initial state as
\begin{equation}|\Psi_2(0)\rangle=|+\rangle\otimes\hat{U}_2^\dag|\{0_k\}\rangle.\label{psi2}
\end{equation}It is noted that this state is different from Eq. (\ref{psi1}), under which it has been shown that the dynamics of the SBM without the RWA exhibits the coherent-incoherent transition. The merit of choosing this state as the initial state is that it takes the form $|\Psi(0)\rangle=\hat{U}_2\hat{U}_1|\Psi_2(0)\rangle=|+_x,\{0_k\}\rangle$ in the $\hat{H}_\text{eff}$ representation. Therefore, besides the zero excitation, only the single-excitation subspace where the bound state is formed is involved in the dynamics. In the same manner as the above RWA case, we can calculate $P_z(t)=\sqrt{2}\text{Re}[h(t)]$, where $h(t)$ satisfies
\begin{equation}\dot{h}(t)+i\eta\Delta h(t)+\int_0^t f'(t-\tau)h(\tau)d\tau=0,
\end{equation}with the initial condition being $h(0)={1\over\sqrt{2}}$ and the kernel function $f'(t-\tau)\equiv{1\over 4\pi}\int_0^{\infty}J'(\omega)e^{-i\omega (t-\tau)}d\omega$ connecting to the renormalized spectral density $J'(\omega)=\sum_{k}\nu^2_{k}\delta(\omega-\omega_k)$.

\begin{figure}[tbp]
\centering
\includegraphics[width = 1 \columnwidth]{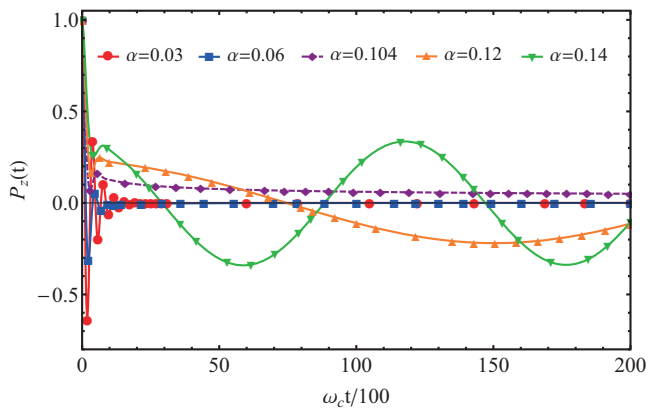}
\caption{(Color online) The nonequilibrium dynamics of $P_z(t)$ of the SBM without the RWA under the initial condition given by Eq. (\ref{psi2}). Here the parameters $s=0.7$ and $\Delta=0.02\omega_c$ have been used. The critical point can be evaluated numerically at $\alpha_\text{C}=0.104$. } \label{dyn2}
\end{figure}

Figure \ref{dyn2} portrays $P_z(t)$ under the initial condition (\ref{psi2}) when the RWA is relaxed. We can see that the similar behavior as Fig. \ref{dyn1} is present. When $\alpha$ is small, the dynamics shows complete decoherence, with the quantum coherence decaying to zero. With the increasing $\alpha$ to certain value, a finite steady $P_z(t)$ can be obtained asymptotically. With the further increasing of $\alpha$, $P_z(t)$ shows decoherence inhibition with the quantum coherence tending to lossless oscillation.

\subsection{QPT without RWA}
In the similar manner as the case under RWA, we expect that such anomalous decoherence is caused by the intrinsic QPT occurring explicitly in the conventional delocalized phase regime.
After the similar procedure as in Sec. \ref{bdt} and neglecting temporarily the constant term $C$ in Eq. (\ref{rwa22}), we can determine that a bound state $\left\vert \varphi_1 ^{\prime }\right\rangle =d_{0}\left\vert
+,\{0\}\right\rangle +\sum_{k}d_{k}\left\vert -,\{1\}_{k}\right\rangle$ with the eigenvalue $E_1$ satisfying
\begin{equation}
y(E_1)\equiv\frac{\eta\Delta}{2}-\sum_{k}\frac{\nu_{k}^{2}}{\omega_{k}-(E_1+\frac{\eta\Delta}{2})}=E_1
\label{add4}
\end{equation}
can be formed for $\hat{H}_\text{eff}$. This equation permits a real root in the regime $E_1\leq-\eta\Delta /2$ if and only if $y(-\eta\Delta /2)\leq-\eta\Delta /2$. Accompanying the formation of a bound state, the ground state is changed from $|\varphi_0'\rangle\equiv|-,{0_k}\rangle$ to $\left\vert \varphi_1 ^{\prime }\right\rangle $. Recovering back the neglected term $C$, we get the ground-state energy as
\begin{equation} \label{grode}
E_g=\left\{ \begin{aligned}
         &-{\eta\Delta\over 2}-C,\ \ \ \ \alpha<\alpha_{_\text{C}} \\
         &E_1-C,\ \ \ \ \alpha>\alpha_{_\text{C}}
                          \end{aligned} \right..
\end{equation}where the critical point $\alpha_{_\text{C}}$ can be determined by solving equation $y(-\eta\Delta /2)=-\eta\Delta /2$. Physically, such a sudden change of the ground-state structure signals clearly the occurrence of QPT in the system.

It is noted that the neglected higher-order perturbation term $\hat{H}_2'$ gives no contribution to the QPT because it is zero, i.e., $\langle \varphi_i'|H_2'|\varphi_j'\rangle=0~(i,j=0,1)$, in the two eigenbases. It means that the neglected term $\hat{H}_2'$ has no impact on such level-crossing-caused QPT. This in turn validates our approximation.

\begin{figure}[h]
\centering
\includegraphics[width = 1 \columnwidth]{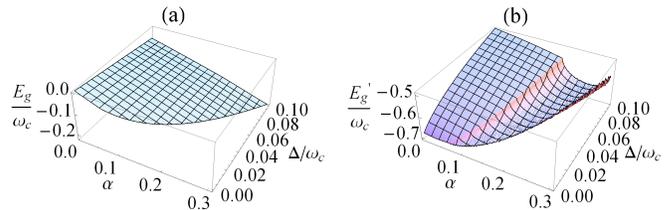}
\caption{(Color online) (a) Ground-state energy $E_g$ and (b) its first derivative $E_g'={\partial E_g\over \partial \alpha}$ as a function of $\alpha$ and $\Delta$.} \label{pas}
\end{figure}

Remembering we are working in the conventional delocalized phase regime, where $\eta$ takes a finite value, we now verify the QPT by studying the ground-state energy of the sub-Ohmic SBM. Taking $s=0.7$ as an example, we plot in Figs. \ref{pas}(a) and \ref{pas}(b), respectively, the ground-state energy and its first derivative to the coupling constant according to Eq. (\ref{grode}). We can see that $E_g$ is continuous, but ${dE_g\over d\alpha}$ shows a discontinuity at the critical point $\alpha_{_\text{C}}$ where the bound state is formed. It manifests clearly that there is another QPT existing in the delocalized phase regime.

From the analysis above, we can see that the QPT induced by the formation of the bound state has a profound impact on the nonequilibrium dynamics of the TLS in the conventional delocalized phase regime. It induces a dynamical transition from complete decoherence to decoherence suppression for the initial states where only the single-excitation subspace is involved. On the other hand, a widely studied case is the nonequilibrium dynamics of the TLS when the reservoir is initially in a vacuum state, where the spin dynamics shows a transition from damped coherent oscillation to incoherent relaxation, i.e., the so-called coherent-incoherent transition, in the delocalized phase regime. Therefore, it is reasonable to conjecture that the coherent-incoherent transition is also a dynamical consequence of this QPT on the initial vacuum state of the reservoir. This has been proved analytically for the Ohmic spectral density in Ref. \cite{Tong2011} that $\alpha_{_\text{C}}$ of the QPT matches well with the point of the coherent-incoherent transition. Thus, we can conclude that both of the transitions from complete decoherence to decoherence suppression and from damped coherent oscillation to incoherent relaxation are actually the different dynamical consequences of the same intrinsic QPT on different initial states.

\section{Conclusion}\label{sum}
In conclusion, going beyond the Born-Markovian approximation, we have studied the decoherence dynamics of the TLS in the SBM both with and without RWA. When the RWA is used, we reveal analytically that a QPT induced by the formation of a bound state in the single-excitation subspace occurs. This QPT causes the anomalous decoherence phenomenon, i.e., the decoherence inhibited with the increasing of the coupling strength. When the RWA is relaxed, using the perturbation approach to neglect the high-order interaction terms in a unitarily transformed Hamiltonian, we have shown that the similar anomalous decoherence induced by the intrinsic QPT still exists in the conventional delocalized phase regime. The approximation is justified by the fact that we are working in the weak-coupling (i.e., the delocalized phase) regime and in the zero-temperature case, where the high-order excitations are negligible. On the other hand, one also can verify that the neglected terms give no contribution to the QPT we obtained, which in turn validates our approximation. This dynamical behavior is compatible with the coherent-incoherent transition which happens to the state where the reservoir is initially in vacuum. It conjectures that the coherent-incoherent transition reported in the literature is essentially caused by the intrinsic QPT of the SBM. Our purely analytical results when the RWA is relaxed agree qualitatively with the results obtained under the path-integral Monte Carlo methods \cite{Kast2013}. Our results also suggest a control way to beat the effect of decoherence by engineering the spectrum of the reservoirs to approach the non-Markovian regime and to form the bound state of the whole system. This can be readily realized in the newly emerged field, i.e., reservoir engineering \cite{Tong2006,Porras2008,Barreiro2011,Murch2012}, for controlling the quantum system by tailoring its coupling to the reservoirs.

\section*{ACKNOWLEDGEMENTS}
This work is supported by the Fundamental Research Funds for the Central Universities, by the NSF of China (Grants No. 11175072, No. 11174115, and No. 10934008), and by the National Research Foundation and Ministry of Education, Singapore (Grant No. WBS: R-710-000-008-271).

\end{document}